\documentclass[
    ,final            
  ]
  {aipproc}

\layoutstyle{6x9}

\begin{document}

\title{Inclusive Inelastic Electron Scattering from Nuclei}


\author{Nadia Fomin}{
  address={University of Virginia, Charlottesville, VA}
}
\begin{abstract}
Inclusive  electron  scattering  from  nuclei  at  large  x and  $Q^2$ 
is  the  result  of  a reaction  mechanism  that  includes  both  
quasi--elastic  scattering  from nucleons and  deep  
inelastic  scattering  from  the  quark consitituents  of  the  
nucleons.
 Data in this regime can be used to study a wide variety of 
topics,  including the extraction of nuclear  momentum distributions,
 the influence of final state interactions and the  approach to $y$-scaling,
  the strength of nucleon-nucleon correlations, and the  approach to
  $x$- scaling, to name a few.  Selected results from the recent experiment E02-019 at the
  Thomas Jefferson National Accelerator Facility will be shown and
their relevance discussed.
\end{abstract}

\maketitle


\section{Scaling in the quasielastic regime}
The quasielastic cross section is analyzed in the Plane Wave Impulse
Approximation (PWIA), where the dominant process is assumed to be
scattering from an individual nucleon, whose motion is treated
relativistically.  
The inclusive cross section for this process is given by ~\cite{CiofidegliAtti:1990rw}
\begin{equation}
\label{qe_cs}
\frac{d^2\sigma}{d\Omega dE}=\sum ^A_{N=1}{\int {dE} \int
  {d^3p}}\:S(p,E) \sigma _{eN} \:\delta(E_0+\nu +\sqrt{M^{*2}_{A-1}+(\vec{p}+\vec{q})^2}) 
\end{equation}
where $S(p,E)$ is the nucleon spectral function, $\sigma_{eN}$ is the
off-shell electron-nucleon cross section.  Following the steps of
~\cite{CiofidegliAtti:1990rw}, we can simplify and factorize this
expression to give us:
\begin{equation}
\label{qe_factored}
\frac{d^2\sigma}{d\Omega dE}=F(y,q) \frac{1}{Z\sigma
  _{ep}+N\sigma_{en}}\frac{\textbf{q}}{\sqrt{M^2+(\textbf{p}+\textbf{q})^2}}.
\end{equation}
F(y,q) is the scaling function defined as
\begin{equation}
F(y,q)=2\pi\:\int ^{\infty}_{E_s^{min}}\:dE\int ^{\infty}_{y1(E^{min}_s)} dp \: p \:S(p,E)
\label{fy_def}
\end{equation}
 where $y$ is the longitudinal momentum of the
struck nucleon, determined by the argument of the $\delta$-function in
Eq.~\ref{qe_cs}.  Expression
~\ref{qe_factored} allows us to extract the response function F(y,q) from the
measured cross section.
\begin{figure}[h!]
\label{sigma_scaling}
  \includegraphics[width=0.95\textwidth]{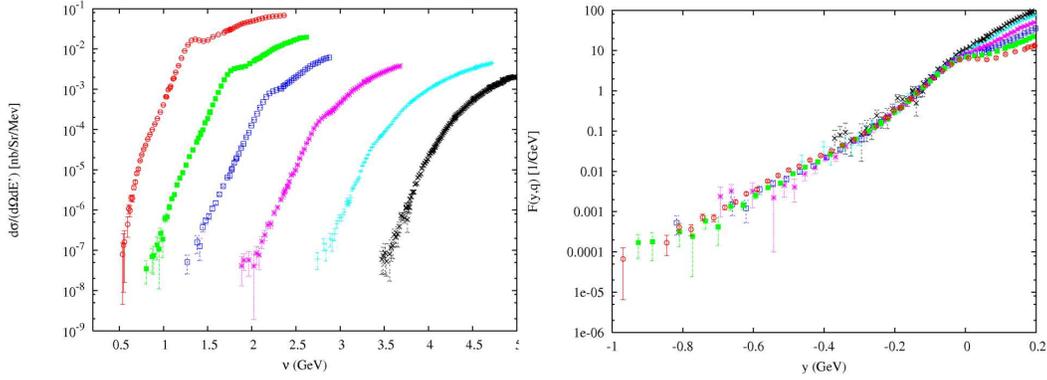}
  \caption{Left: $^3$He Cross sections for 6 scattering angles (18$^{\circ}$-50$^{\circ}$) as a function of
  $\nu$, energy loss, at 5.766GeV (JLab experiment E02-019). Right: The scaling functions extracted
  from the above cross sections as a function of $y$, the longitudinal
  momentum of the struck nucleon.  $Q^2$ range covered is
  2.5-7.4 at the quasielastic peak.
 Scaling is observed for negative values of $y$,
  corresponding to the $x>1$ region, where quasielastic contributions
  are dominant.}
\end{figure}
Scaling of the reduced cross sections (F(y,q)) in $y$ has been
observed ~\cite{Day:1993md, Arrington:1998ps}  in the
quasi-elastic region (Example: Fig.~\ref{sigma_scaling}).  This means that the
reduced cross section, instead of being a function of energy and
momentum transfers, instead depends on only one variable, in this
case $y$, and the reduced cross sections for different kinematics all
lie on one scaling curve.

 The scaling function can be related to the nucleon momentum
 distribution, but there are a few obstacles.  First, there's a
 binding correction that comes from the possibility of having the
 recoil nucleus be in an excited state.  Also, the PWIA approach does not
take Final State Interactions (FSIs) into account, whose contributions
are largest at low values of $Q^2$ and large negative $y$'s and enough
 to change
 the predicted approach to scaling.
 And finally, in order to extract a momentum distribution, one needs to be able to subtract or reasonably
neglect any contributions from inelastic processes which requires
either a cross-section model or data taken at kinematics where
inelastic processes don't contribute.  

\section{Short Range Correlations}
In Fig.~\ref{sigma_scaling}, we observe the struck nucleons of very high
momenta.  The repulsive NN force imparts high
momentum to the nucleons as the interaction distance between the
nucleons becomes smaller than the average inter-nucleon spacing.
The momentum of fast nucleons is balanced by the correlated
nucleon(s), rather than the rest of the nucleons.

The ideal regime for studying SRCs is at $x$>1 and $Q^2$>1, where
scattering from low-momentum nucleons is suppressed and the energy
transfer is higher than the kinetic energies of the correlated
nucleons.  Here, the mean field contribution is negligible, and the
inclusive cross section can be approximated with ~\cite{Frankfurt:1993sp}
\begin{equation}
\sigma_A(x,Q^2)=\sum ^A _{j=2}\: \frac{a_k(A)}{j}\sigma_j(x,Q^2) 
\end{equation}
where $\sigma_A(x,Q^2)$ is the electron-nucleus cross section,
$\sigma_j(x,Q^2)$ is the electron-$j$-nucleon-correlation cross section,
and $a_j(A)$ is proportional to the probability of finding a nucleon
in a $j$-nucleon correlation.
\begin{figure}[]
\label{xem_src}
  \includegraphics[width=0.95\textwidth]{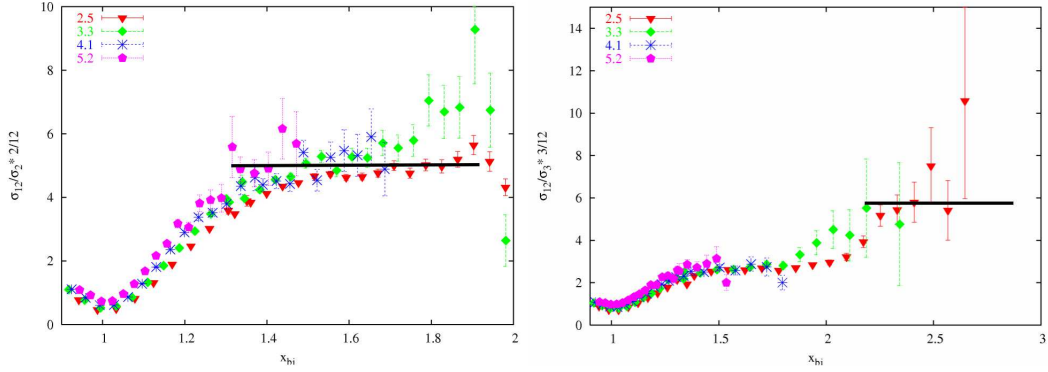}
  \caption{SRC ratios for $^{12}$C.  The left plot shows the ratio to $^2$H
  and the right plot shows the ratio to $^3$He.  The ratio for
  1.4<$x$<2 is proportional to the abundance of the NN correlations in
  Carbon, and the ratio for x>2.4 is proportional to the abundance of
  3N correlations.}
\end{figure}
Using $^2$H and $^3$He cross sections as well as a theoretical
calculation for the fraction of nucleons in each of those that are in
a 2- or 3-N correlation, one can obtain SRCs for heavier targets,
which show the abundance of correlations.  For example, for $^{12}$C,
shown in Fig.~\ref{xem_src} has a $\approx$20$\%$ and $\approx$0.6$\%$ probability of a
2N and 3N SRC, respectively, which is in agreement with previous
measurements ~\cite{Egiyan:2005hs}.
\section{Inelastic Scattering}
As one can see from Fig. ~\ref{sigma_scaling}, $y$-scaling fails
for $y$>0, where inelastic processes dominate.  In this region, the cross
section is described in terms of the nuclear structure functions:
\begin{equation}
\frac{d^2\sigma}{d\Omega
  dE}=\sigma_{mott}[W_2^A(\textbf{q},\nu)+2W_1^A(\textbf{q},\nu) \tan^2
  (\theta /2)]
\end{equation}
In the limit of high energy loss and momentum transfer in the DIS regime, the structure
functions simplify to functions of $x$ alone.  When one examines the
structure functions with the Nachtmann variable $\xi$ ~\cite{Nachtmann:1973mr}, which  extends the scaling to lower
values of $Q^2$, one expects to
see the same scaling behavior as is seen for $x$ in the same kinematic
region.  $\xi$ reduces to $x$ as $Q^2\rightarrow
\infty$ ($\xi=2x/(1+\sqrt{1+4M^2x^2/Q^2}$).

\begin{figure}[h!]
\label{x_xsi_scaling}
  \includegraphics[width=0.95\textwidth]{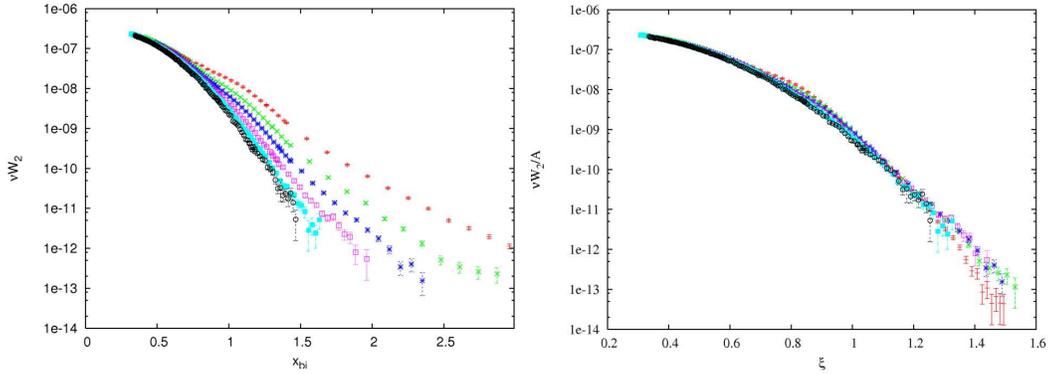}
  \caption{Structure function $\nu W_2^A$ for $^{12}C$ as a function of
  $x$ (left) and $\xi$ (right).  Scaling is only observed for low
  values of $x$, but for all values of $\xi$, rather than just in the
  inelastic region. }
\end{figure}
Our analysis of the inelastic cross section is done through the study
of the $\nu W_2^A$ structure function, which can be extracted from the
measured cross section in the following way:
\begin{equation}
\nu W_2^A=\frac{d^2\sigma}{d\Omega dE}\cdot \frac{\nu}{\sigma
  _{mott}[1+2\: \tan^2(\theta /s)\frac{1+\nu^2/Q^2}{1+R}]}\:\rm{,}
\end{equation}
where $R$ is the ratio of the longitudinal cross section to the
transverse ~\cite{Bosted:1992fy}.

Fig.~\ref{x_xsi_scaling} shows that the $\nu W_2^A$ structure function
scales in $x$ in the DIS region.  Since the $\xi$ variable is
analogous to $x$, we expect $\nu W_2$ to scale in $\xi$ as well, in
this region. However, scaling is observed for all values of $\xi$, including the quasielastic region.

  The first explanation ~\cite{Benhar:1995rh} of this observation involves expanding $\xi$ as a function of $y$,
suggesting that the scaling we see in the quasielastic region is the same
as the $y$-scaling analysis yields, with the presence of FSIs masked
by the relationship between the two variables.  Another explanation
~\cite{Day:2003eb} states that the observed scaling is purely
accidental and is just a result of the inelastic contribution falling
off at the same rate as the quasielastic contribution rises.

A final explanation for this observation is that it might be due to local duality
~\cite{Bloom:1971ye}.  Bloom and Gilman 
observed that the structure function has the same $Q^2$ behavior as the resonance form
factors.  In fact, the scaling curve is recovered by averaging over
the resonance peaks.
\section{Conclusion}
Despite the fact that the contributions from different 
processes measured in inclusive electron scattering are difficult to separate, it still offers a
wealth of information.  With a good sample of data over a variety of
targets and kinematics, we can study nucleon momentum distributions,
SRCs, scaling, $Q^2$ behavior of the $\nu W_2^A$
structure functions, as well as other physics.  


\bibliographystyle{aipproc}   

\bibliography{fomin_bib}

\IfFileExists{\jobname.bbl}{}
 {\typeout{}
  \typeout{******************************************}
  \typeout{** Please run "bibtex \jobname" to optain}
  \typeout{** the bibliography and then re-run LaTeX}
  \typeout{** twice to fix the references!}
  \typeout{******************************************}
  \typeout{}
 }

\end{document}